\documentclass{PoS}

\usepackage{url}

\title{HELP-ing Extragalactic Surveys : The Herschel Extragalactic Legacy Project and The Coming of Age of Multi-Wavelength Astrophysics}

\ShortTitle{HELP-ing Extragalactic Surveys : The Herschel Extragalactic Legacy Project}

\author{\speaker{Mattia Vaccari}\\
           Department of Physics \& Astronomy, University of the Western Cape, Cape Town, South Africa\\
           INAF - Istituto di Radioastronomia, via Gobetti 101, 40129 Bologna, Italy\\
        E-mail: \email{mattia@mattiavaccari.net}}

\abstract{How did galaxies form and evolve? This is one of the most challenging questions in astronomy today. Answering it requires a careful combination of observational and theoretical work to reliably determine the observed properties of cosmic bodies over large portions of the distant Universe on the one hand, and accurately model the physical processes driving their evolution on the other. Most importantly, it requires bringing together disparate multi-wavelength and multi-resolution spectro-photometric datasets in an homogeneous and well-characterized manner so that they are suitable for a rigorous statistical analysis. The Herschel Extragalactic Legacy Project (HELP) funded by the EC FP7 SPACE program aims to achieve this goal by combining the expertise of optical, infrared and radio astronomers to provide a multi-wavelength database for the distant Universe as an accessible value-added resource for the astronomical community. It will do so by bringing together multi-wavelength datasets covering the 1000 deg$^2$ mapped by Herschel extragalactic surveys and thus creating a joint lasting legacy from several ambitious sky surveys.}

\FullConference{SALT Science Conference 2015\\
		 1-5 June 2015\\
		 Stellenbosch Institute of Advanced Study, South Africa}

\begin{document}



\maketitle

\section{The HELP project \& its science objectives}\label{intro.sec}

How did galaxies form and evolve? This is one of the most challenging questions in astronomy today. Although astronomers now have a good understanding of the background cosmology and of the formation of the large-scale structure of the dark matter \cite{Springel2005,Vogelsberger2014}, the complex astrophysics that leads to the variety and numbers of galaxies observed within dark matter halos is still very poorly understood. Anecdotal clues to these questions can be found through focussed studies of individual galaxies. However, the fundamental requirement for rigorous testing of any theories of galaxy formation and evolution is a complete statistical audit or census of the stellar content and star formation rates of galaxies in the Universe at different times and as a function of the mass of the dark matter halos that host them. This audit requires many elements. We need unbiased maps of large volumes of the Universe made with telescopes that probe the different wavelengths at which the different physical processes of interest manifest themselves. We need catalogues of the galaxies contained within these maps with photometry estimated uniformly from field-to-field, from telescope-to-telescope and from wavelength-to-wavelength. We need to understand the probability of a galaxy of given properties appearing in our data sets. We need the machinery to bring together these various data sets and calculate the value-added physical data of primary interest, e.g. the distances (or redshifts), stellar masses, star-formation rates and the actual number densities of the different galaxy populations.

Since the advent of large-format detector arrays at optical and infrared wavelengths, many astronomical facilities have been undertaking ambitious programmes to observe large areas of the sky and study galaxy evolution in the distant Universe. ESA's Herschel mission had a unique place in this context, probing the obscured star-formation history of the Universe, i.e. roughly 50\% of all star formation activity, through observations at far-infrared and sub-millimeter wavelengths. Herschel extragalactic surveys were a major goal of Herschel and accounted for around 10\% of the observing time available during Herschel's 3.5 year science mission. However, the full exploitation of this dataset is complicated by Herschel's large instrumental beam size (6-37" at 70-500 $\mu$m).

The Herschel Extragalactic Legacy Project (HELP, PI : Seb Oliver, University of Sussex, \url{http://herschel.sussex.ac.uk}) brings together several of the teams that have been undertaking ambitious coordinated multi-wavelength digital sky surveys to study large volumes of the distant Universe over the past decade. These observational projects are mature enough that we are now able to undertake the necessary homogenization and thus provide the first representative and comprehensive census of the galaxy populations in the distant Universe. HELP was therefore funded by the European Commission FP7-SPACE-2013-1 Scheme (Grant Agreement 607254) to produce over the 2014-2017 4-year funding period a database for the distant Universe as an accessible value-added resource for the astronomical community. HELP was conceived to assemble the ancillary data and develop the tools necessary to fully capitalize on Herschel's legacy and to enable astronomers not directly involved with the mission to fully exploit the Herschel dataset and its multi-wavelength counterparts.  We intend to provide a vast resource for studying the distant Universe, similar to the SDSS for the nearby Universe, as a lasting legacy of several major ground-based and space-based surveys and a solid foundation for future space missions and ground-based observatory projects.
\section{The Herschel satellite \& its mission}\label{herschel.sec}

Galaxies emit electromagnetic radiation over a very wide wavelength range, but most of it is absorbed by the Earth's atmosphere and thus cannot be studied from the ground. With the development of satellite missions, however, astronomers have gradually been able to explore galaxy emission over the full electromagnetic spectrum, from $\gamma$-rays to radio waves, in an uninterrupted manner. Comparing the properties of galaxies observed at different wavelengths have thus enabled substantial progress in the understanding of the physical processes driving their formation and evolution.

The Herschel satellite \cite{Pilbratt2010} was developed by ESA and carried out its 3.5 year science mission between 2009 and 2013. Herschel was the first 4-m class space telescope and the first far-infrared and sub-millimeter telescope sensitive enough to detect galaxies in the distant Universe, vastly improving the state of observations in this poorly explored band. Herschel thus signaled the completion of an ambitious program jointly carried out by ESA and NASA to map galaxy evolution across most of the life of the Universe at all wavelengths. In particular, the Herschel imaging instruments SPIRE \cite{Griffin2010} and PACS \cite{Poglitsch2010} fully constrain the peak of the far-infrared and sub-millimeter background with six photometric channels covering the 70-500 $\mu$m wavelength range. Herschel thus allows us to thoroughly investigate the sources of the infrared background radiation and characterize their total obscured star formation as a function of cosmic time through systematic observing programs such as HerMES \cite{Oliver2012} and H-ATLAS \cite{Eales2010a}.

However, the large size of the Herschel beam (6-37" across the 70-500 $\mu$m wavelength range) means that the objects that can be clearly seen as individual sources only make up a small fraction of the cosmic infrared background. In order to unlock the full information available in Herschel maps one must therefore develop new methods to estimate the short-wavelength sources most likely to contribute to emission in Herschel bands in a statistically rigorous manner. HELP therefore brings together leading scientists working on Herschel extragalactic surveys with experts in multi-wavelength data reduction, analysis and homogenization to make sure that we can effectively build upon existing best practices to deliver the highest-quality multi-wavelength datasets and selection functions.

\section{HELP data products \& Herschel's scientific legacy}\label{df.sec}

Many of the outstanding problems in galaxy evolution require a multi-wavelength approach to properly account for different physical processes over the full life-cycle of galaxies and black holes. Modeling tools to account for the Spectral Energy Distribution of galaxies have become sufficiently advanced that mis-matched photometry (i.e. photometry at different wavelengths which samples different areas of a galaxy) may be the dominant source of error. Thus in all wavebands it is now common to produce "aperture-matched" catalog. In this approach the source detection is carried out on either a single image or, for the greatest sensitivity, on a combined image, but then the photometry is carried out on all images with a common aperture or a common model for the galaxy light profile. However, statistical studies of galaxy populations also require a detailed understanding and modeling of the selection processes in the basic data products and the derived properties. These "selection functions" are seldom available for individual datasets and rarely, if ever, published for derived physical properties, but are crucial to estimate source number densities in a reliable manner \cite{Vaccari2010,Eales2010b,Burgarella2013,Marchetti2015}.

The starting point for our project will be the multi-wavelength images and catalogs that have been produced and publicly released by our team as well as by others based on more than a decade of concerted observational projects \cite{Lonsdale2003,Mauduit2012,Jarvis2013}. Our first task is to homogenize these data, so that photometry and calibration is consistent from catalog to catalog, field to field, and from wavelength to wavelength, building upon the framework developed as part of the Spitzer Data Fusion (\cite{Vaccari2010,Marchetti2015}, \url{http://www.mattiavaccari.net/df/}). The benchmark for this will be that we can determine accurate photometric redshifts, or distances, adopt consistent galaxy spectral templates and characterize the physical properties of the galaxies \cite{RowanRobinson2008,RowanRobinson2013}. For this purpose we will also assemble available optical spectroscopy from, e.g. SDSS, 2dFGRS, GAMA, PRIMUS, VVDS, zCOSMOS and provide homogeneous reliability flags for both spectroscopic and photometric redshifts. We will thus produce a key input catalogs for future spectroscopic surveys with, e.g. WEAVE and WAVES.

\begin{figure*}
\centering
\includegraphics[width=10.5cm]{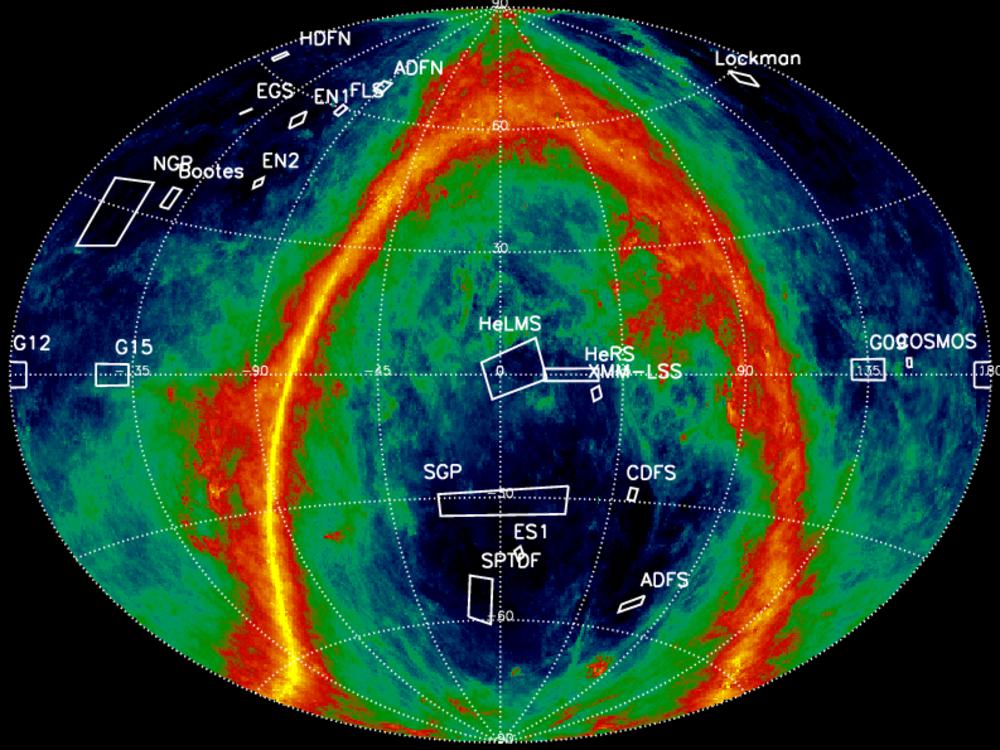}
\caption{The HELP fields overlaid on the IRAS/COBE dust maps by \cite{Schlegel1998} in ecliptic coordinates.}
\label{sky-help.fig}
\end{figure*}

\begin{figure*}
\centering
\includegraphics[width=10.5cm]{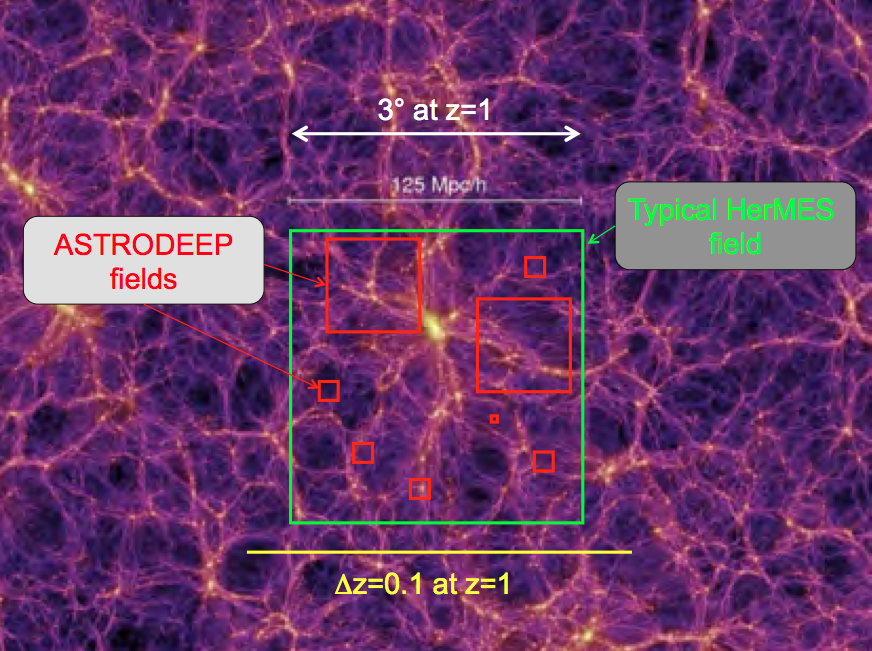}
\caption{A slice of the dark matter in the Universe today from the Millennium Simulation of the Universe by \cite{Springel2005}. Overlaid are various scale markers, including the footprints of the AstroDeep fields and of one of $3 \times 3$ deg$^2$ HerMES fields, illustrating how much of the progenitors of these structures they would sample at $z = 1$. The smallest H-ATLAS field is about twice the size of this entire image. HELP covers an area 15 times as large as the whole image.}
\label{footprint.fig}
\end{figure*}

HELP will thus produce and publicly release multi-wavelength datasets (catalogs and images) and selection functions as well as physical parameters for individual galaxies over the 1,000 deg$^2$ covered by the Herschel extragalactic surveys (See Figure~\ref{sky-help.fig}). The data products as well as the techniques and tools that we will produce will enable astronomers to fully capitalize on datasets provided by Herschel as well as other surveys (See Table~\ref{surveys.tab}. This will extend the kinds of scientific investigations made possible a decade ago in the nearby Universe by the SDSS into the early Universe and provide a lasting legacy for surveys and facilities in the future. In so doing, HELP will provide a complementary view to the AstroDeep project focusing on five small sky areas totaling 2 deg$^2$ and thus sampling fainter galaxies but not the full range of environments, rare objects and short-lived phases in galaxy evolution (see Figure~\ref{footprint.fig}).
\section{Conclusion}\label{conclusion.sec}
%
HELP (Herschel Extragalactic Legacy Project, \url{http://herschel.sussex.ac.uk}) will produce an unrivaled public database of multi-wavelength sky images, catalogs of individually detected sources and their physical properties as well as related selection functions over the 1,000 deg$^2$ of the extragalactic sky covered by the Herschel satellite. It will bring together in a concerted manner space-based and ground-based wide-area galaxy surveys for the first time, enabling highly-accurate statistical studies of galaxy properties and their evolution with cosmic time and providing a lasting legacy for the astronomical community to mine for decades to come and a template for future space science data curation projects in the petascale and exascale era.
%
\begin{table*}
\centering
\begin{tiny}
\begin{tabular}{llll}
\hline\noalign{\smallskip}
Wavelength & Telescope / Instrument & Observing Band & Survey Project  \\
\noalign{\smallskip}\hline\noalign{\smallskip}
Ultraviolet & GALEX & FUV \& NUV & DIS, MIS, AIS\\
Visible & PS1, SDSS, DECAM, VST, CFHT, INTWFC & $ugrizy$ & PS1, SDSS, DES, ATLAS, KIDS, VOICE, INTWFS \\
Near-Infrared & 2MASS, UKIRT, VISTA & $ZYJHK$ & UKIDSS, VIDEO, VIKING, VHS \\
Mid-Infrared & IRAC, WISE & 3.6/4.5/5.8/8.0/12.0 $\mu$m & WISE, SWIRE, SERVS, S-COSMOS, SpUDS, SPLASH \\
Far-Infrared & WISE, MIPS, PACS & 22/24/70/100/160 $\mu$m & WISE, SWIRE, PEP, HerMES\\
(Sub-)Millimeter & SPIRE, SCUBA, SCUBA2, LABOCA, AzTEC, ACT, SPT & 250/350/500/850 $\mu$m & HerMES, SHADES, S2LS, ACT, SPT \\
Radio & ATCA, GMRT, LOFAR, MeerKAT, JVLA & 0.6, 1.4, 3, 5 GHz & ATLAS, LOFAR, WODAN, MIGHTEE, VLASS \\
\noalign{\smallskip}\hline\noalign{\smallskip}
\end{tabular}
\end{tiny}
\caption{A Selection of Relevant Multi-Wavelength Survey Projects within HELP Fields.}
\label{surveys.tab}
\end{table*}
\section*{Acknowledgements}
HELP is funded by the EC REA (FP7-SPACE-2013-1 GA 607254 - Herschel Extragalactic Legacy Project).
Mattia Vaccari is also supported by the Square Kilometre Array South Africa Project, the South African NRF
and DST (DST/CON 0134/2014) and Italy's MAECI (PGR GA ZA14GR02 - Mapping the Universe on the Pathway to SKA).
%
%
%

\end{document}